\documentclass[pra,preprint]{revtex4}
\usepackage{amssymb,amsmath,amsthm,graphicx}
\begin{document}

\newcommand{\A}{\ensuremath{\mathcal{A}}}
\newcommand{\B}{\ensuremath{\mathcal{B}}}
\newcommand{\AB}{\ensuremath{\mathcal{AB}}}
\newcommand{\Sys}{\ensuremath{\mathcal{S}}}
\newcommand{\Env}{\ensuremath{\mathcal{E}}}
\newcommand{\I}{\ensuremath{\mathcal{I}}}
\newcommand{\Ise}{\ensuremath{\I_{\Sys:\Env}}}
\newcommand{\Isen}[1]{\ensuremath{\I_{\Sys:\Env_{#1}}}}
\newcommand{\Tr}{\mathrm{Tr}}
\newcommand{\ket}[1]{\ensuremath{\left|#1\right\rangle}}
\newcommand{\bra}[1]{\ensuremath{\left\langle#1\right|}}
\newcommand{\braket}[2]{\ensuremath{\left\langle#1|#2\right\rangle}}
\newcommand{\expect}[1]{\ensuremath{\left\langle#1\right\rangle}}
\newcommand{\braopket}[3]{\ensuremath{\left\langle#1\right|#2\left|#3\right\rangle}}
\newcommand{\z}{\ensuremath{\vec{z}}}
\newcommand{\V}{\ensuremath{\mathcal{V}}}
\newcommand{\Ibar}{\ensuremath{\bar{\I}}}
\newcommand{\Hh}{\ensuremath{\mathcal{H}}}
\newcommand{\Hs}{\ensuremath{\Hh_{\Sys}}}
\newcommand{\He}[1]{\ensuremath{\Hh_{\Env_{#1}}}}
\newcommand{\Hse}{\ensuremath{\Hh_{\Sys\Env}}}
\newcommand{\Ham}{\ensuremath{\bf \mathrm{H}}}
\renewcommand{\tensor}{\ensuremath{\otimes}}
\newcommand{\Tensor}{\ensuremath{\bigotimes}}
\newcommand{\Envset}[1]{\ensuremath{\Env_{\left\{#1\right\}}}}
\newcommand{\diff}{\ensuremath{\mathrm{d}\!}}

\newtheorem{lemma}{Lemma}
\newtheorem{definition}{Definition}

\title{A simple example of ``Quantum Darwinism'': Redundant information storage in many-spin environments}
\date{March 15, 2004}
\author{Robin Blume-Kohout}
\email{rbk@socrates.berkeley.edu}
\affiliation{Los Alamos National Laboratory}
\author{Wojciech H. Zurek}
\affiliation{Los Alamos National Laboratory}
\begin{abstract}
As quantum information science approaches the goal of constructing quantum computers, understanding loss of information through
decoherence becomes increasingly important.  The information about a system that can be obtained from its environment can facilitate quantum control and error correction.  Moreover, observers gain most of their information indirectly, by monitoring (primarily photon) environments of the ``objects of interest.''  Exactly \emph{how} this information is inscribed in the environment is essential for the emergence of ``the classical'' from the quantum substrate.  In this paper, we examine how many-qubit (or many-spin) environments can store information about a single system.  The information lost to the environment can be stored redundantly, or it can be encoded in entangled modes of the environment.  We go on to show that randomly chosen states of the environment almost always encode the information so that an observer must capture a majority of the environment to deduce the system's state.  Conversely, in the states produced by a typical decoherence process, information about a particular observable of the system is stored redundantly.  This selective proliferation of ``the fittest information'' (known as Quantum Darwinism) plays a key role in choosing the preferred, effectively classical observables of macroscopic systems.  The developing appreciation that the environment functions not just as a garbage dump, but as a communication channel, is extending our understanding of the environment's role in the quantum-classical transition beyond the traditional paradigm of decoherence.
\end{abstract}
\maketitle

\section{Introduction}
In recent years, the field of \emph{quantum information science} has grown tremendously.  Quantum cryptography is rapidly becoming the domain of engineers more than of physicists; quantum optics and nanotechnology are making quantum systems more and more directly accessible; and quantum computers are seen not as a pipe dream, but as the next century's inevitable technology.

In this context, many scientific questions of fundamental interest have donned new hats as pressing technological issues.  The transition from quantum to classical, a subject that used to be the field of philosophers and quantum foundationalists, is suddenly a critical area of research -- where is it, and how does it happen?  If we are to design quantum hardware, then one of the primary failure modes is precisely its transition to classical behavior.  The process by which this transition occurs is decoherence, and decoherence is accordingly more interesting than ever today.

That quantum environments interact with quantum systems and, by doing so, decohere them, is by now widely appreciated.  Until recently, models of decoherence usually (from a conceptual standpoint) stopped at that point; the challenge was to identify (a) what the environment was doing to the system, and (b) what was the resulting evolution of $\rho_{\Sys}$, the state of the system.  The environment obviously gained information about the system -- effectively measuring it -- but what happened to that information once it had been sucked away was largely irrelevant.  The standard ansatz of decoherence analysis reflects this: from the state $\ket\Psi$ of the entire universe, one obtains the system's state $\rho_{\Sys}$ by tracing out the environment $\Env$.  Once $\Env$ is traced over, it vanishes from the scene.

Recently, the environment has started to become interesting again \cite{ZurekPTRSA98, ZurekADP00, ZurekRMP03, Ollivier03, Zurek03}.  The most obvious feature that distinguishes the classical realm from the quantum is \emph{objectivity}.  Unknown classical states can be found out without being disturbed, so they are said to ``exist objectively.''  This is, of course, not the case for quantum states.  Thus, as it has often been emphasized by Asher Peres \cite{Peres95,FuchsPT00}, the nature of a quantum state is a much more elusive thing.  In quantum systems (at least isolated ones) such states cannot be regarded as existing objectively. As was pointed out some time ago \cite{ZurekPTP93}, however, observers usually obtain information about a system by measuring its environment.  Therefore, the manner in which the environment stores and transmits information is of great interest.  In a sense, the environment is not just a wastebasket (where quantum information gets thrown away), but rather a communication channel through which observers find out about their universe.  In this ``environment as a witness'' view of the transition from quantum to classical, objectivity is a consequence of the redundancy \cite{ZurekADP00,ZurekRMP03,Ollivier03,Zurek03} with which information is inscribed in $\Env$.  The environment has also become interesting because a greater understanding of the entire decoherence process can facilitate the protection of fragile quantum information.  Quantum control, for instance, is often based on ``second hand'' information obtained from the environment.  On the other hand, it's much easier to recover all of the lost quantum information when a single copy is stored in a local environment than when trillions of copies have been recorded throughout the universe.

In this paper, we skip almost completely over the traditional approaches to decoherence, and examine instead some very general properties of quantum states.  We ignore the processes that produce states in which the environment has information about the system, analyzing instead the way in which that information is recorded in the environment.  Our ``toy'' universe consists of a collection of qubits (or spin-$\frac12$ particles), one of which is the system $\Sys$.  We begin by asking a very simple question: When the whole environment has got some information $\I$ about the system, how much of that $\I$ can be found out by measuring a single qubit, or two qubits, or $m$ qubits?  We ask this question about arbitrary states of the universe, and then we ask it about a particular sub-ensemble of states.

We conclude that our experience of the ``real classical universe'' is consistent with states that record certain information redundantly at the expense of complementary information.  Such states appear naturally in the process of decoherence.  Redundantly recorded observables of the system are in effect objective; they can be independently deduced by many observers from the state of $\Env$ that was responsible for decoherence \cite{ZurekADP00,ZurekRMP03,Ollivier03,Zurek03}.  Finally, we examine the concept of redundant information storage directly, and relate this quantitative measure of redundancy to our more qualitative tools for analyzing joint system-environment states.

\section{Laying out the playing field}
In any discussion of decoherence, we have to start by defining a system (\Sys) and an environment (\Env).  In order to go further
and examine the redundancy of information storage, we also need to partition $\Env$ into sub-environments $\Env_1\dots\Env_N$.  For this
entire discussion, then, our universe will be an $(N+1)$-qubit Hilbert space $\Hh = \Hs \tensor \Tensor_{i=1}^{N} \He{i}$, where both
$\Hs$ and all the $\He{i}$ are single qubits (or spin-$\frac12$'s).

Our paradigm for understanding decoherence rests on the idea that environments decohere the system by measuring it, and that in measuring the system the environments come to have information about the system.  Of the several quantities we could choose to measure \emph{how much} information an environment $\Env$ has about the system, we choose {\bf quantum mutual information} (QMI), which is the amount of entropy produced by destroying correlations between $\Sys$ and $\Env$:
\begin{equation}
\Ise \equiv H(\Sys) + H(\Env) - H(\Sys\Env) \label{MIDef}
\end{equation}
While other appealing measures of information exist \cite{Ollivier03}, the QMI is mathematically elegant and generally easy to compute.  We note here a few properties of $\I$ for future reference.
\begin{itemize}
\item $\Ise = 0$ if and only if $\rho_{\Sys\Env} = \rho_{\Sys} \tensor \rho_{\Env}$ (i.e., the subsystems are uncorrelated).
\item When $\rho_{\Sys\Env}$ is a pure state, $\Ise = 2H(\Sys)$, since $H(\Env) = H(\Sys)$ and $H(\Sys\Env) = 0$.
\item When $\rho_{\Sys\Env}$ is a pure state, and $\Env$ is partitioned into $\Env_1$ and $\Env_2$,  $\Isen{1} + \Isen{2} = \Ise$.  This follows simply from expanding $\I$ using Eq. \ref{MIDef}, and the symmetry of bipartite entanglement for pure states.  Note that this formula is \emph{not} true for mixed $\rho_{\Sys\Env}$, or for divisions of $\Env$ into more than two sub-environments!
\item For pure states of $\Sys\Env$, $\Ise$ is also an entanglement measure.  This connects our work with an existing body of work on multipartite entanglement, particularly in randomly or dynamically produced states, (see, e.g., \cite{KendonJMO02,KendonPRA02}), and on the ``entangling power'' of dynamical maps (see e.g. \cite{ScottJPA03,ScottPRA04}, and refs. 24-35 in \cite{ScottPRA04}.  In this paper, however, we intentionally avoid the complicated field of entanglement by treating mutual information only as a measure of correlation.
\end{itemize}

Within this framework, the overall question we examine is: {\bf How do individual environments' information (about $\Sys$) compare to the \emph{whole} environment's information?}.  We'll assume that no particular environment is ``special''; rather, we'll examine the information possessed by a typical environment or collection thereof.  It should be noted that this is \emph{not} the same as assuming that all environments are identical; we allow environments to have more or less information about the system, but we obtain a statistical ``typical environment'' by averaging.  This assumption reduces our overall question to a more definite one: {\bf How does the mutual information $\I$ between the system and a typical sub-environment $\Envset{m}$ of size $m$ depend on $m$?}

\begin{figure}
\includegraphics[width=0.9\columnwidth]{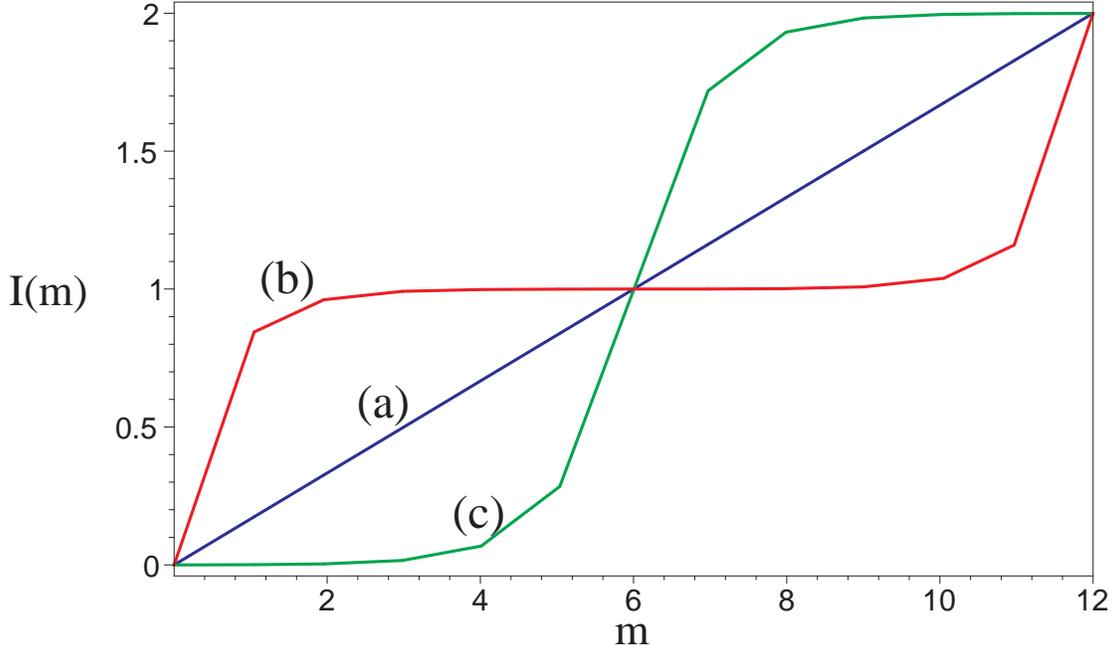}
\caption{The three basic profiles for partial information plots ($\I$ vs. $m$).  Curve (a) shows the behavior of \emph{independent} environments; curve (b) shows \emph{redundantly} stored information; and curve (c) illustrates the case where information is \emph{encoded} in multiple environments.}
\label{ThreePIPs}
\end{figure}

Our primary tool for answering this question is the \emph{partial information plot} or PIP (see Fig. \ref{ThreePIPs}), which plots $\Isen{\{m\}}$ against $m$.  Since there are many sub-environments $\Envset{m}$ containing $m$ individual environments, the partial information plot is not a curve, but a scatter-plot or histogram.  To obtain a smooth curve, we average over all $\Envset{m}$ to obtain $\bar{\I}(m)$.  In later sections, we'll discuss the advantages and disadvantages of this averaging scheme (as opposed to, e.g., taking the average of $m$ over all $\Envset{m}$ with information $\I$).  Before examining specific PIP's, we prove a simple but important lemma.

\begin{lemma} 
When $\rho_{\Sys\Env}$ is a pure state, $\bar\I(m)$ is antisymmetric around $m=\frac{N}{2}$. \label{SymmetryLemma}
\end{lemma}

\begin{proof} As noted previously, when $\Env$ is partioned into $\Envset{m}$ and $\Envset{N-m}$, then $\Isen{\{m\}} + \Isen{\{N-m\}} = \Ise$.  We obtain $\bar\I(m)$ by averaging $\Isen{\{m\}}$ over all sub-environments of size $m$, and for each $\Envset{m}$ there exists a unique complement $\Envset{N-m}$.  Thus
\begin{eqnarray*}
\bar{\I}(m) + \bar{\I}(N-m) &=& \binom{N}{m}^{-1}\sum_{\{m\}}{\Isen{\{m\}}} \\
 			    && +\binom{N}{N-m}^{-1}\sum_{\{N-m\}}{\Isen{\{N-m\}}}\\
&=& \binom{N}{m}\sum_{\{m\}}{\left( \Isen{\{m\}} + \Isen{\{N-m\}} \right)}\\
&=& \binom{N}{m}\sum_{\{m\}}{\Ise}\\
&=& \Ise
\end{eqnarray*}
Since $\bar\I(m)+\bar\I(N-m) = \Ise$, $\bar\I(m)$ must be antisymmetric around $m=\frac{N}{2}$; further, 
$\bar\I\left(\frac{N}{2}\right) = \frac12\Ise = H(\Sys)$.\end{proof}

From Lemma \ref{SymmetryLemma} and the obvious condition that $\bar\I(m)$ must be non-decreasing, the PIP can take three basic shapes (Fig. \ref{ThreePIPs}). The simplest profile (a) is one where $I(m) \propto m$ -- that is, each environment provides unique and independent information, so that the total information gained is simply proportional to the number of environments examined.  Alternatively (b), information may be \emph{redundantly} stored; $\bar\I(m)$ increases rapidly at first, then plateaus at $\bar\I \sim H(\Sys)$.  Finally (c), information about the system may be \emph{encoded}, so that $\bar\I(m)$ remains close to $0$, then increases rapidly around $m \sim \frac{N}{2}$.  As we shall see, each of these cases is realizeable and physically relevant.

\section{Information storage for randomly chosen states\label{RandomStatesSection}}

In order to have a baseline to compare other ensembles of states against, we start by examining the typical PIP for the ensemble of \emph{all} states (the uniform ensemble).  To perform this average requires a measure; we use the unique invariant measure induced by the Haar measure over the unitary group $U(2^{N+1})$.  This simplifies the process, since averaging over all possible sub-environments $\Envset{m}$ with $m$ components is neatly subsumed in the average over all states.  

In fact, while the calculation can be done numerically, there exists an analytic solution.  Page \cite{PagePRL93} conjectured a formula for the mean entropy (averaged over Haar measure) of a subsystem, which was later proven by Sen \cite{SenPRL96} and others \cite{FoongPRL94,SanchezRuizPRE95}.  Page's formula for the mean entropy $\bar{H}(m,n)$ of an $m$-dimensional subsystem of an $mn$-dimensional system (where $m\leq n$) is
\begin{equation}
\bar{H}(m,n) = \sum_{k=n+1}^{mn}{\frac{1}{k}} - \frac{m-1}{2n}.
\end{equation}
We use this formula to calculate the mean entropy $\bar{H}_{k;N}$ of a $k$-qubit ($2^k$-dimensional) subsystem of an $N$-qubit ($2^N$-dimensional) universe, obtaining an expression in terms of the digamma function $\Psi(n)$:
\begin{equation}
\bar{H}_{k;N} = \Psi(2^N+1)-\Psi(2^{(N-k)}+1) - 2^{k-N-1}(2^k-1)
\end{equation}
The average mutual information between a 1-qubit system and an $m$-qubit sub-environment, all within an $N+1$-qubit universe, is
\begin{equation}
\overline{\Isen{\{m\}}} = \bar{H}_{1;N+1} + \bar{H}_{m;N+1} - \bar{H}_{m+1;N+1} \label{HaarMI}.
\end{equation}

\begin{figure}
\includegraphics[width=0.9\columnwidth]{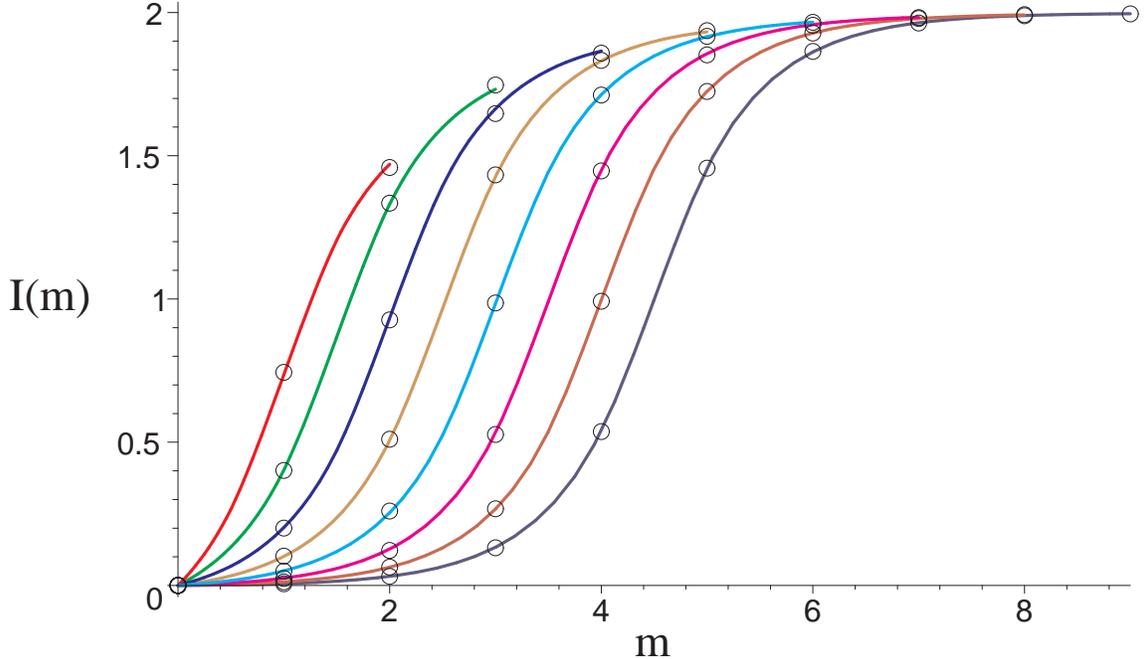}
\caption{Partial information plots (both $\I$ and $m$ are measured in bits), averaged over \emph{all} states, for environments of 2 to 9 qubits.  The circles represent numerical computations, while the solid lines represent theory (analytically continued to non-integer $m$).  The plots indicate that information about the system is encoded in entangled states of the environment; nearly $N/2$ individual environments must be captured to yield more than a pittance of information.  At $m = N/2$, where $\bar\I(m)$ climbs most rapidly, its slope is almost exactly 1.}
\label{HaarMIplots}
\end{figure}

PIPs computed from Equation \ref{HaarMI} are shown in Fig. \ref{HaarMIplots}, along with brute-force numerical calculations which agree extremely well with Equation \ref{HaarMI}.  The plots demonstrate that for even a modestly large ($N \simeq 9$) environment, the vast majority of randomly-chosen states are ``coding states'', which encode the information about the system so that it cannot be accessed without capturing at least $\sim N/2$ environments.  Such states display no redundancy whatsoever; when three or more observers attempt to divvy up the environment so that each has the same amount of information, each observer learns nothing.  

Since the single-qubit system's entropy is limited to 1 bit, $\Ise \leq 2$; nonetheless, for environments larger than about 5 qubits, almost every state achieves the bound.  An additional calculation also confirms the indication in Figs. \ref{ThreePIPs} and \ref{HaarMIplots}, that the slope of $\bar\I(m)$ at $m=N/2$ is almost exactly 1. Thus, for large $N$, almost the entire 2 bits of available information are gained between $m=\frac{N}{2}-1$ and $m=\frac{N}{2}+1$.  As we shall see in the next section, these results don't imply that information is encoded this way in \emph{every} state; rather, the vast majority of states are ``coding states.''

\section{Information storage for product environments \label{ProductEnvironmentSection}}

The results of the previous section indicate that the vast majority of states are non-redundant coding states, in which information is recorded so that multiple observers can never independently measure the same system.  Our everyday experience, however, contradicts this result.  By intercepting a tiny fraction of the photons (or phonons) which scatter off (or are emitted by) another human being, we obtain a great deal of information about their appearance (or speech); moreover, hundreds or thousands of other human beings obtain essentially the same information!  We are thus led to conjecture \cite{ZurekADP00,ZurekRMP03} that the ensemble of system-environment states produced by natural processes is substantially different from the ensemble of all states.

Physical processes that produce decoherence typically have a tensor-product structure.  The system interacts independently with each environment, and the environments interact only weakly with each other.  We thus consider interaction Hamiltonians of the form $\Ham(\Sys,\Env) = \sum_{i}{\Ham(\Sys,\Env_{i})}$.  If the initial states of the environment are uniformly distributed over the Haar measure of $\He{}$, however, then the action of \emph{any} set of unitaries will yield a uniformly distributed set of states.  In order to get a ensemble which is not uniform, we consider only initial product states: $\ket{\Psi_{\Sys\Env}} = \ket{\psi_{\Sys}} \tensor \Tensor_i{\ket{\psi_{\Env_{i}}}}$.  Under these assumptions, the ``universe'' will evolve into
states of the form
\begin{equation}
\ket{\Psi_{\Sys\Env}} = \alpha\ket{0}_{\Sys}\Tensor_i{\ket{\psi_i}_{\Env_{i}}}
			+ \beta\ket{1}_{\Sys}\Tensor_i{\ket{\psi'_i}_{\Env_{i}}}, \label{StateEnsemble}
\end{equation}
where $\ket{0}$ and $\ket{1}$ are used, without loss of generality, to indicate the pointer basis of the system \cite{ZurekRMP03,FuchsPT00}.  This model is obviously close to the one discussed (using a different approach, and asking different questions) in \cite{Ollivier03}.  Subtly different ensembles of such states can be constructed, depending on the measure used for $\ket{\psi}_{\Sys}$ and $\ket{\psi}_{\Env_i}$, but the most important feature of these states is the structure of correlations.  The system is strongly entangled with the overall environment, and potentially with some of the sub-environments, but the environments are never entangled with each other.  Strong classical correlations, however, typically exist between all subsystems.  Essentially, all correlations are mediated through $\Sys$.

Such states turn out to be quite amenable to mutual information calculations.  Because of the correlation structure, every relevant reduced density matrix is at most rank-2 -- that is, a state of a virtual qubit.  In order to obtain mutual informations, we compute the density matrices for (a) a reduced universe consisting of $\Sys$ and a set $\Env_m = \{i_1\ldots i_m\}$ of environments; (b) just $\Sys$; and (c) just $\Env_m$.  Using the definition $\gamma_i = \braket{\psi_i}{\psi'_i}$, we obtain:
\begin{eqnarray}
\rho_{\Sys\Env_m} & = & \left(\begin{array}{cc} \left|\alpha\right|^2 & \alpha^*\beta\left(\prod\limits_{i\notin\Env_m}{\gamma_i}\right) \\
					\alpha\beta^*\left(\prod\limits_{i\notin\Env_m}{\gamma^*_i}\right) & |\beta|^2 \end{array}\right) \label{rho_se}\\
\rho_{\Sys} & = & \left(\begin{array}{cc} |\alpha|^2 & \alpha^*\beta\left(\prod\limits_{i}{\gamma_i}\right) \\
					\alpha\beta^*\left(\prod\limits_{i}\right) & |\beta|^2 \end{array}\right) \label{rho_s}\\
\rho_{\Env_m} & = & \left(\begin{array}{cc} |\alpha|^2 & \alpha^*\beta\left(\prod\limits_{i\in\Env_m}{\gamma_i}\right) \\
					\alpha\beta^*\left(\prod\limits_{i\in\Env_m}{\gamma^*_i}\right) & |\beta|^2 \end{array}\right) \label{rho_e} \\
\end{eqnarray}
We characterize each density matrix of the form
\begin{equation}
\rho = \left(\begin{array}{cc} x & \sqrt{x(1-x)}\gamma\\ \sqrt{x(1-x)}\gamma^* & 1-x \end{array}\right)
\end{equation}
by a \emph{base purity} $P_0 = x^2 + (1-x)^2$ and a \emph{additive decoherence factor} $d = -log\left(|\gamma|^2\right)$.  The idea is that $P_0$ represents
the susceptibility of $\rho$ to decoherence in the diagonal basis, while $d$ represents the amount of decoherence ($d=0$ indicates a pure state, while
total decoherence is indicated by $d=\infty$).  Conveniently, it's also appropriate to think of $d$ as representing \emph{distinguishability}, since it measures how nearly orthogonal $\ket{\psi_i}$ and $\ket{\psi'_i}$ are.  In terms of $P_0$ and $d$, the Von Neumann entropy $H = -\Tr(\rho\log\rho)$ is:
\begin{eqnarray}
H(P_0,d) &=& \ln2 - \frac12\left[(1+z)\ln(1+z) + (1-z)\ln(1-z)\right] \nonumber \\
\mbox{where\ }&& z = \sqrt{1-2(1-P_0)\left(1-e^{-d}\right)} 
\end{eqnarray}
The key advantage to this treatment is that the $d$-factors are additive.  If we define $d_i = -log{|\gamma_i|^2}$, then the decoherence factors
for $\Sys\Env_m$, $\Env_m$, and $\Sys$ are
\begin{eqnarray}
d_{\Sys\Env_m} & = & \sum_{i \notin \Env_m}{d_i} \\
d_{\Env_m} & = & \sum_{i \in \Env_m}{d_i} \\
d_{\Sys\Env_m} & = & \sum_{\mbox{all } i}{d_i} \\
		 & = & d_{\Sys\Env_m} + d_{\Env_m}, \nonumber
\end{eqnarray}
while the base purities are all the same. Thus, in order to compute the mutual information between $\Sys$ and an environment $\Env_m$, we need only
know $P_0$, $d_{\Sys}$ (a measure of the information lost to the \emph{full} environment), and $d_{\Env_m}$ (a measure of the amount of information
held by $\Env_m$).  Given these, the mutual information is simply
\begin{equation}
\Isen{\{m\}} = H(P_0,d_{\Sys}) + H(P_0,d_{\Env_m}) - H(P_0,d_{\Sys}-d_{\Env_m}),
\end{equation}
and in cases where $P_0$ and $d_{\Sys}$ are fixed, we can simply consider $\I(d_{\Env_m})$.  For the rest of this paper, we'll assume that $P_0=\frac12$ -- that is, that the initial state has one full bit of information to be measured.  Our investigations indicate that most of our results are almost identical for other values of $P_0$, however.

\begin{figure}
\includegraphics[width=0.9\columnwidth]{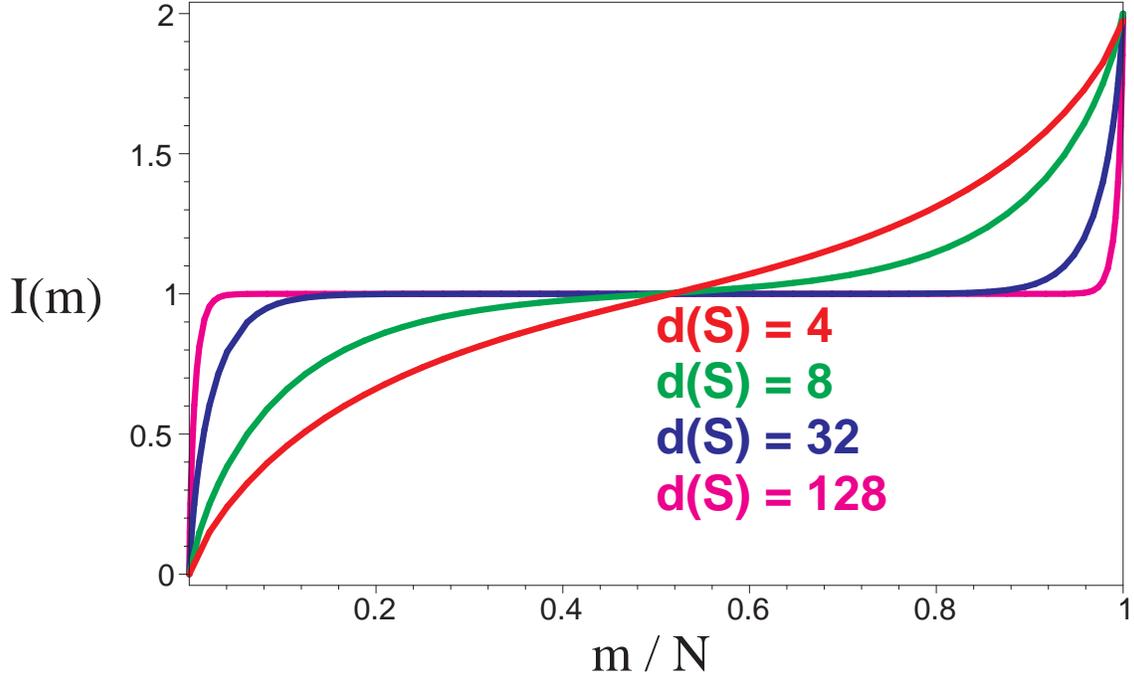}
\caption{Partial information plots ($\I(m)$) for unimodal distributions of environment $d$-factors.  $\I(m)$ is plotted against the fraction $m/N$ of environments captured.  As the total $d_{\Sys}$ increases from 2 to 8 to 32 to 128, the plots become more sharply inflected near $m=0$ and $m=N$, while the central plateau at $\I = 1$ bit grows flatter.  In the limit as $d_{\Sys} \longrightarrow \infty$, $\I(m)$ jumps immediately to 1 bit at $m=0^+$, then jumps again to 2 bits at $m=N^-$.}
\label{UnimodalPIPs}
\end{figure}

The simplest product-environment state is one where the states $\ket{\psi_i}$ and $\ket{\psi'_i}$ are equally distinguishable for each $i$; that is, $d_i = d_0,\ \forall\ i$.  In this case, $\Isen{\{m\}} = \I(md_0)$, and no averaging over distinct sub-environments is necessary.  In Fig. \ref{UnimodalPIPs}, we plot such PIPs for various values of $d_{\Sys}$.  Note that the only free parameter is the total decoherence $d_{\Sys}$.  For small $d_{\Sys}$, the total mutual information between system and environment is substantially less than $2 H(\Sys)$, and $\Isen{\{m\}}$ increases almost linearly with $m$.  As $d_{\Sys}$ increases, $\Ise$ approaches $2 H(\Sys)$, and the $\I(m)$ curve becomes more sharply curved, indicating increased redundancy.  Perfect redundancy -- where a sub-environment has all the classical information that the entire environment has -- is achieved only for a GHZ-type state with $d_{\Sys} = \infty$.

Of course, we can imagine a state in which the system is fully decohered ($d_{\Sys} = \infty$), but not all sub-environments possess perfect information.  More generally, the $d_i$ may be unequal, so that some environments have more information than others.  In this case, to obtain a valid PIP we must actually perform an average over sub-environments.  Given a distribution function $f_1(d)$ for the $d_i$, we can obtain a distribution function $f_m(d)$ for the total decoherence of $m$ randomly chosen environments.  The PIP is then given by 
\begin{equation}
\bar\I(m) = \int{f_m(d)\I(d) \diff d}. \label{PIPIntegral}
\end{equation}
In general, both obtaining $f_m(d)$ and computing the given integral are nontrivial.  We therefore present two sample cases.

\subsection*{Case 1: A single state}
Of particular interest is the case where the state $\ket\Psi$ is known and has a finite number $N$ of sub-environments.  In this case, the $d_i$ form a finite collection, and so both $f_1(d)$ and $f_m(d)$ are discrete probability distributions.  A general technique for large $N$ is to treat the process of sub-environment selection as a random walk: $f_1(d)$ is characterized by a mean value $\bar{d}$ and a width $\Delta d$.  The Central Limit Theorem then predicts that as $m$ gets large, regardless of what $f_0(d)$ actually is, $f_m(d)$ is very well approximated by a Gaussian distribution.  In a conventional random walk, $\bar{d}_m$ is simply $m\bar{d}$, and $\Delta d_m = \sqrt{m}\Delta d$.  Our example is slightly different, in that there exists a finite collection of $d_i$, and when $m=N$, each of them must have been picked.  A good analogy is a random walk which \emph{must}, after $N$ steps, arrive at a known endpoint.  Thus, while $\bar{d}_m$ still increases linearly with $m$, the width of $f_m(d)$ increases to a maximum at $m = N/2$, then falls to $0$ again at $m=N$.  The precise result is:
\begin{eqnarray}
\bar{d}_m &=& m\bar{d} \\
\Delta d_m &=& \sqrt{m\left(1-\frac{m-1}{N-1}\right)} \Delta d
\end{eqnarray}
Using this ansatz to approximate $f_m(d)$ is quite effective when $N$ is large; unfortunately, integrating Equation \ref{PIPIntegral} is then analytically intractable, so its usefulness is restricted to numerical computations.

A simple example of this sort of analysis which \emph{is} tractable occurs when there are only two kinds of environments: ``good'' environments with $d=d_0$, and ``useless'' environments with $d=0$.  Such an environment can be characterized by three quantities: the total number $N$ of environments; the number $n_u$ of useful environments, and the distinguishability $d_0$ of each useful environment.  When $m$ environments are randomly selected, the probability of choosing $m_u \in \left[\mbox{max}(0,m+n_u-N)\ldots\mbox{min}(m,n_u)\right]$ useful environments can be computed with elementary combinatorics:
\begin{eqnarray}
P_m(m_u) &=& \frac{\binom{n_u}{m_u}\binom{N-n_u}{m-m_u}}{\binom{N}{m}} \\
f_m(d) &=& \sum_{m_u}{P_m(m_u) \delta(d-m_ud_0)}.
\end{eqnarray}
We integrate Equation \ref{PIPIntegral} to obtain
\begin{equation}
\bar{\I}(m) = \sum_{k = \mbox{max}(0,m+n_u-N)}^{\mbox{min}(m,n_u)}{\frac{\binom{n_u}{k}\binom{N-n_u}{m-k}}{\binom{N}{m}}\I(kd_0)}
\end{equation}

\begin{figure}
\includegraphics[width=0.45\textwidth]{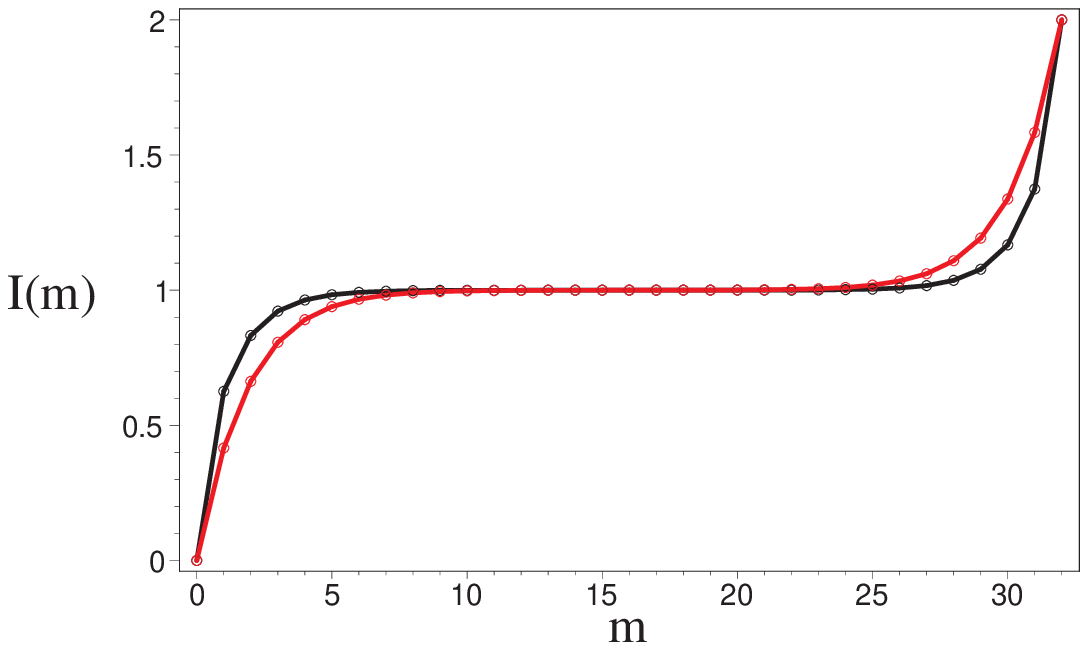}
\includegraphics[width=0.45\textwidth]{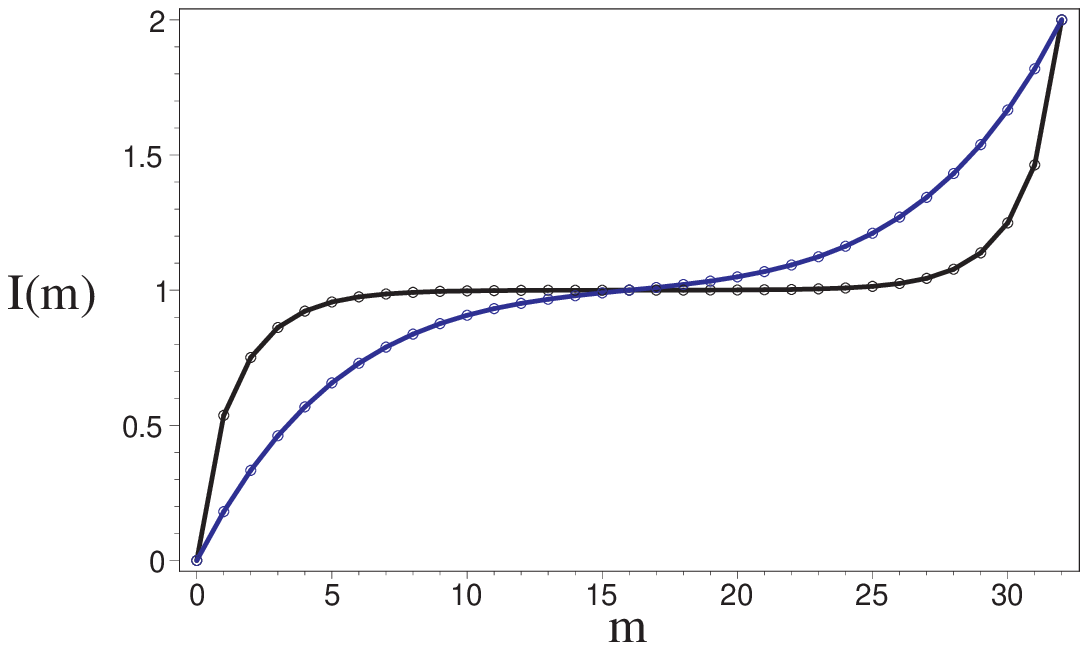}
\caption{Partial information plots ($\I(m)$) for bimodal distributions of environment $d$-factors.  The red curve shows $\bar{\I}(m)$ for an environment where half of the $N=32$ environments have $d=1.5$ and the other half have $d=0$.  The blue curve shows $\bar{\I}(m)$ when 6 environments have $d=3$ and the rest are useless ($d=0$).  The corresponding black curves demonstrate what happens when the equivalent amounts of $d_{\Sys}$ are spread evenly over all 32 environments -- respectively, $d_i=\frac34$ and $d_i=\frac{9}{16}$.  In both cases, there is a smoothing effect, but the smoothing is most pronounced when $d_{\Sys}$ is concentrated in just a few environments.}
\label{BimodalPIPs}
\end{figure}

Fig. \ref{BimodalPIPs} compares exact $\bar{\I}(m)$ plots with the naive approximation $\bar{\I}(m) = \I(n_ud_0/N)$, which achieves the same $d_{\Sys}$ with indistinguishable environments.  In all cases, the bimodal distribution produces a smoother PIP, which rises less quickly and has less of a ``plateau'' in the middle.  This is a simple consequence of the averaging procedure, which effectively convolves $\I(d)$ with a Gaussian, and thus produces a version of the original curve which is greatly smoothed around $m=N/2$, and relatively less smoothed near the ends.

\subsection*{Case 2: An ensemble of states \label{CaseIISection}}
In Section \ref{RandomStatesSection}, we obtained an average PIP for randomly chosen states.  For comparison, we'll now obtain an average PIP for all the product-environment states in Equation \ref{StateEnsemble}.  We select the $\ket{\psi_i}$ and $\ket{\psi'_i}$ randomly from the Haar measure for a single qubit, which is uniform over the Bloch sphere.  A simple calculation shows that $\gamma_i = \left|\braket{\psi_i}{\psi'_i}\right|^2$ is uniformly distributed between 0 and 1, which yields a perfect Poisson distribution for $d_i$ and $d_m$:
\begin{eqnarray}
f_0(d) &=& e^{-d} \\
f_m(d) &=& \frac{d^{m-1} e^{-d}}{(m-1)!}
\end{eqnarray}
Thus armed, we proceed as in Section \ref{RandomStatesSection}, splitting $\Isen{\{m\}}$ into its three component entropies $H(\Sys)$, $H(\Envset{m})$, and $H(\Sys\Envset{m})$, and averaging them independently.  Computing the average entropy of a qubit that has been decohered by $m$ environments turns out to involve a horrendous integral which yields infinite sums of Riemann zeta functions $\zeta(j)$.  In the interest of brevity, we provide only the result here.
\begin{eqnarray}
\bar{H}_m &=& \int{f_m(x)\I(x) d\!x} \nonumber \\
&=& \frac{1}{(m-1)!}\int_0^{\infty}{d\!x x^{m-1} e^{-x}\left[ \ln(2) - \frac{\left(1+e^{-\frac{x}{2}}\right)\ln\left(1+e^{-\frac{x}{2}}\right) + \left(1-e^{-\frac{x}{2}}\right)\ln\left(1-e^{-\frac{x}{2}}\right)}{2}\right]} \nonumber \\
&=& \left(\frac{3^m-2^m}{3^m}\right)\left(\ln(2)-1\right) + \frac{m}{2} - \frac12\sum_{j=2}^{m}{\left(\frac{3^{m+1-j}-2^{m+1-j}}{3^{m+1-j}}\right)\zeta(j)} 
\end{eqnarray}

\begin{figure}
\includegraphics[width=0.45\textwidth]{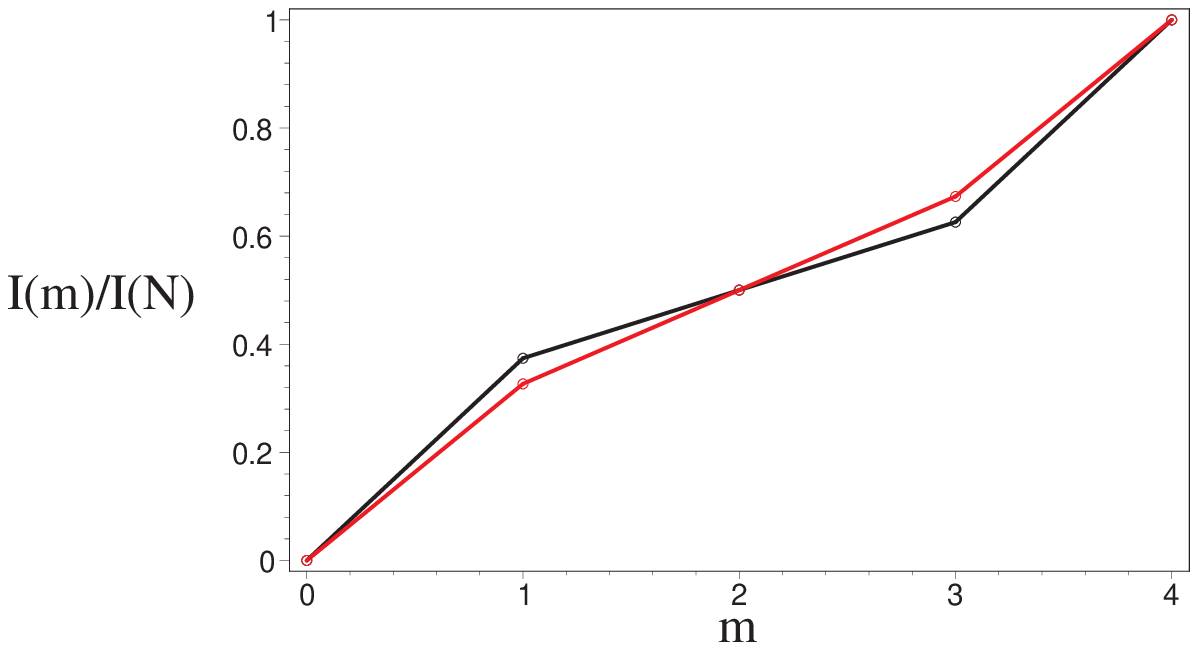}
\includegraphics[width=0.45\textwidth]{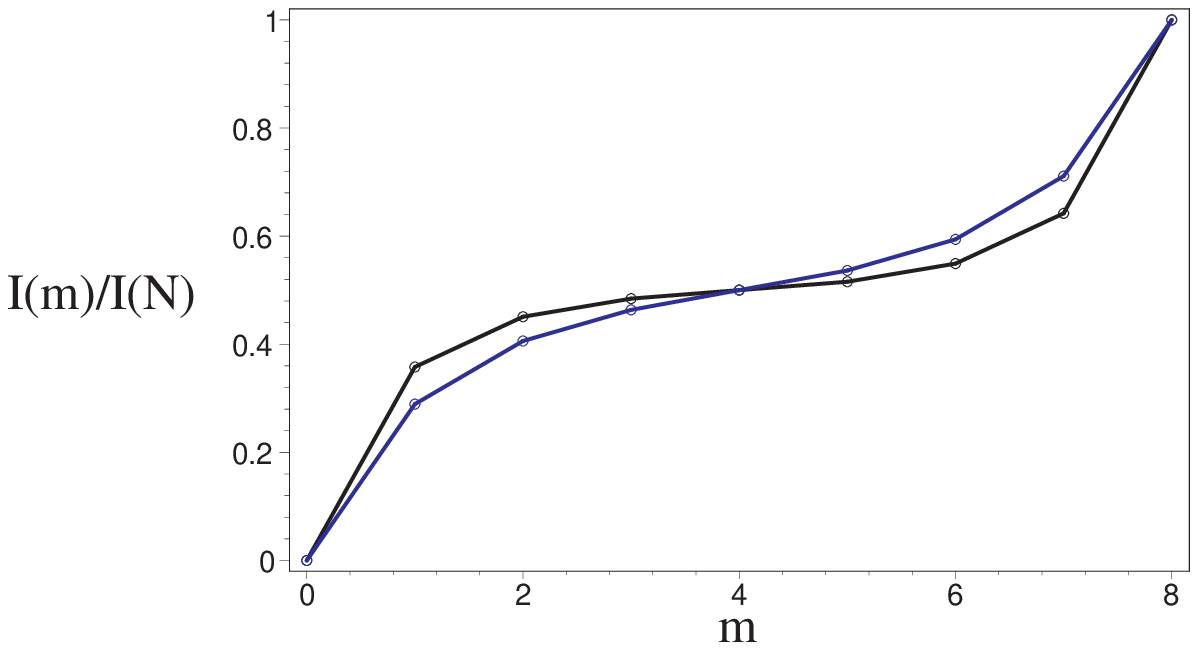}

\includegraphics[width=0.45\textwidth]{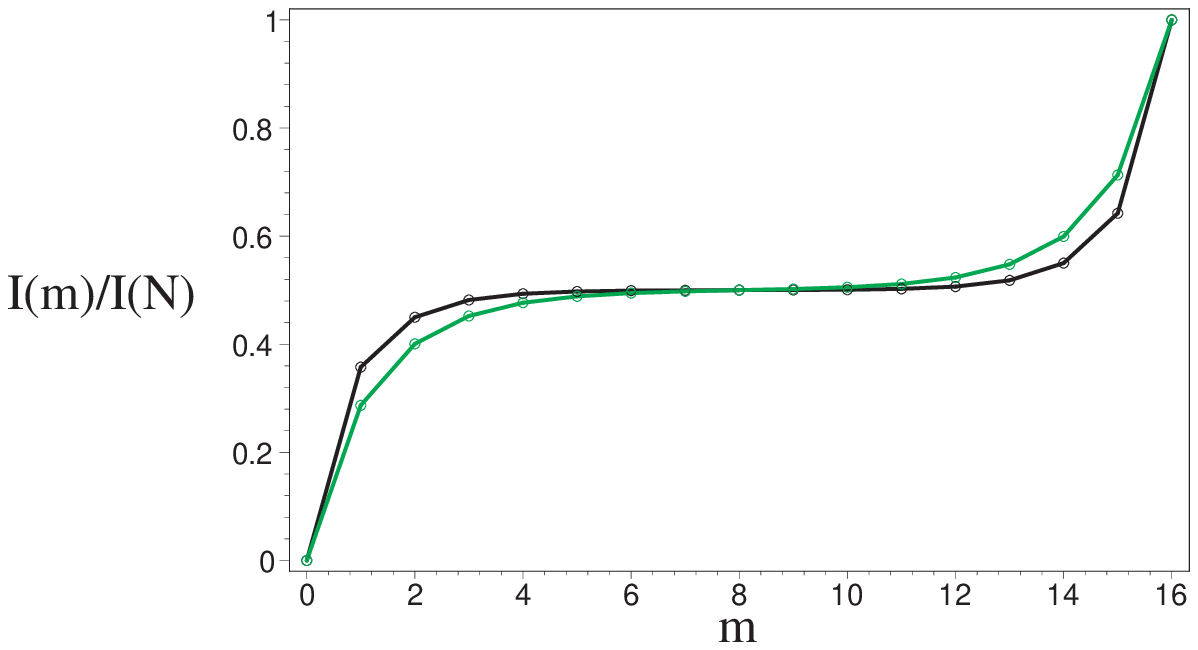}
\includegraphics[width=0.45\textwidth]{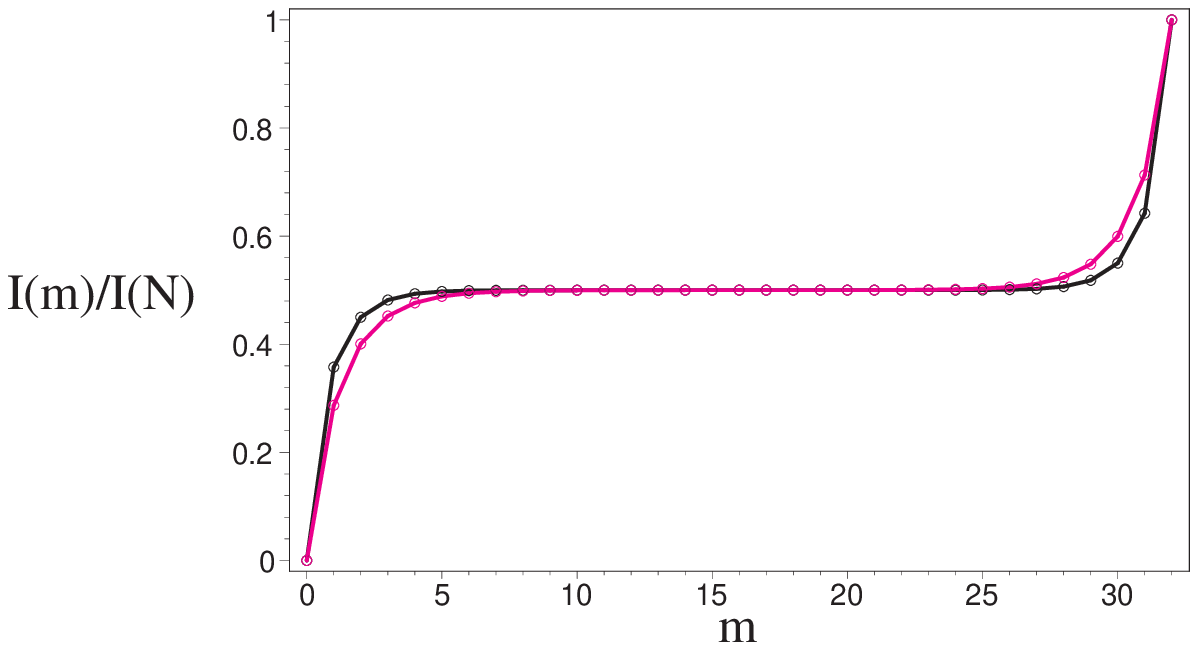}
\caption{Partial information plots ($\bar{\I}(m)$) averaged over the ensemble of all product-environment states.  The four plots represent different numbers $N$ of environments: the red, blue, green, and purple plots correspond respectively to $N=4,8,16,32$.  The associated curves in black show the behavior of unimodal distributions with the same average $\bar{d}_m = m$.  While the ensemble averages are (as expected) less inflected than the unimodal plots, redundant information storage is still quite evident.  The ensemble averages are less smoothed than those obtained from the bimodal distribution (Fig. \ref{BimodalPIPs}), since ensemble average yields a relatively sharply-peaked Poisson distribution for $f_m(d)$.  \emph{NOTE: all plots have been rescaled on the vertical axis by $\bar{\I}(N)$ for ease of comparison}.}
\label{AvgProductPIPs}
\end{figure}

Fig. \ref{AvgProductPIPs} shows averaged product-environment PIPs for various $N$.  These plots contrast sharply with those in Fig. \ref{HaarMIplots}; instead of hovering near $\I=0$, then shooting up to $\I = \Ise$ at $m=N/2$, they rise rapidly to $\I\simeq H(\Sys)$, then plateau until nearly $m=N$.  For the typical product-environment state, a small fraction of the environment yields a great deal of information about the system.  The typical random state, on the other hand, encodes the information so that even a large minority of the sub-environments yields nothing.  Fig. \ref{AvgProductPIPs} also shows that, while typical product-environment states are slightly less generous with their information than the simple unimodal states discussed earlier, the difference is relatively small.  We conclude, then, that product-environment states \emph{as a class} display a much higher level of redundancy in information storage than typical unstructured states.

\section{Redundancy}
The notion of redundancy has been mentioned, but so far not addressed in sufficient detail.  Redundancy was introduced in this context in \cite{ZurekADP00,ZurekRMP03}, and developed into a more precise tool in \cite{Ollivier03}, where the information about $\Sys$ was considered $R$-fold redundant if and only if the environment could be partitioned into $R$ sub-environments, each of which had nearly full information (``nearly full'' meaning that it could supply a fraction $1-\delta$ of the total information).  Reference \cite{Ollivier03} defines such ``$\delta$-redundancy'' using Shannon mutual information between observables of the system $\Sys$ and subsystems of the environment.  We follow the idea of asking about ``nearly all'' information, but depart from \cite{Ollivier03} by using quantum mutual information as our metric (thus choosing the other option from the two possibilities considered in \cite{ZurekADP00,ZurekRMP03}).

Using the quantum mutual information forces us to change our definition of ``nearly full information.''  The metric of information used in \cite{Ollivier03} measure \emph{classical} mutual information -- information about a single observable.  Since it's possible for each of many environments to have the same amount of information about an observable, ``nearly full'' in the context of \cite{Ollivier03} means $\I_{subenvironment} \geq (1-\delta)\I_{total}$.  The quantum mutual information, by contrast, measures information about all possible observables -- qubits A and B can have \emph{two} bits of quantum mutual information if they are entangled, instead of just one.  Since we find in general that, by this metric, the most information about $\Sys$ that can be shared among many environments is $\frac12 \I_{total}$ (see Fig. \ref{ThreePIPs}), we redefine ``nearly full information'' as $\I_{subenvironment} \geq \frac12(1-\delta)\I_{total}$.

We note, however, that if 2 environments each have equal information about \emph{different} parts of the system (this is only possible with systems larger than a single qubit), then each has $\frac12\I_{\Sys\Env}$ mutual information with the system, yet there is no redundancy at all.  In fact, in such a case it's not possible to tell for certain whether the stored information is 1-fold redundant (i.e., not redundant at all), or 2-fold redundant.  In general, if $R$ sub-environments have $\frac12\I_{\Sys\Env}$ information, then the actual redundancy of the information could be anywhere between $R-1$ and $R$.  For this reason, we take the lower bound, and define redundancy as follows:

\begin{definition} When the full environment $\Env$ has $\I_{\Sys\Env}$ mutual information with the system, and can be partitioned into at least $R_{\delta}+1$ sub-environments $\Env_i$ such that each $\Env_i$ has at least $\frac{1-\delta}{2}\I_{\Sys\Env}$ mutual information with the system, then the information is stored with at least $R_{\delta}$-fold redundancy, and we say that $R_{\delta}$ is the redundancy (with fudge-factor $\delta$). \label{RedundancyDefinition}\end{definition}

\section{Implications}

The main implication of the preceding analysis is clear and simple: arbitrary states of a many-part environment \emph{encode} information in entangled modes of the parts, whereas product-environment states of the type likely to be produced by decoherence store the information \emph{redundantly}.  However, a number of more subtle implications deserve attention.

\subsection{Redundancy and $\bar{\I}(m)$ \label{RedundancyImplicationsSection}}
The relationship between redundancy and the $\bar{\I}(m)$ plot requires some elucidation.  If we consider Definition \ref{RedundancyDefinition}, it's apparent that $R_{\delta}$ can be calculated for product-environment states with very little trouble.  For a particular $d_{\Sys}$, let $d_r$ be the minimum value of $d_{\Envset{m}}$ such that $\I(d_{\Sys},d_{\Envset{m}}) = \frac{1-\epsilon}{2}\I_{\Sys\Env}$.  Then any sub-environment $\Envset{m}$ with $d_{\Envset{m}} \geq d_r$ has at least $\frac{1-\epsilon}{2}\I_{\Sys\Env}$ mutual information with the system.  Assuming infinite divisibility of the environment, the number of such sub-environments that can be created by a partition of $\Env$ is precisely $d_{\Env}/d_r$, so the redundancy can be calculated exactly as
\begin{equation}
R = \frac{d_{\Env}}{d_r} -1
\end{equation}
The assumption of infinite divisibility is not really justified, but (a) optimizing the partition exactly is a very tedious combinatoric problem with no closed-form solution, and (b) the $R$ obtained this way is not only a strict upper bound on the exact $R$, but also more useful for some purposes.  For instance, when $d_{\Env} = 10$, $d_r = 1$, and each of 11 environments has $d_i = \frac{10}{11}$, the strict definition of $R$ would force us to partition $\Env$ into 5 sub-environments, each of which has almost twice the required $d$.  The assumption of infinite divisibility essentially allows us to partition $\Env$ into 10 sub-environments containing 1.1 environments each, which (while physically absurd) better reflects the fact that the information is much more than 4-fold redundant.

The simplicity of this analysis begs the question: why examine $\bar{\I}(m)$ in the first place?  Not only does $\bar{\I}(m)$ seem to be irrelevant to redundancy, but in some cases (see the small-$N$ plots in Fig. \ref{AvgProductPIPs}) it can be actively misleading!  The answer is that $\bar{\I}(m)$ can be calculated and analyzed for a wide range of states, whereas the redundancy calculation just given is only simple for the sub-ensemble of product-environment states.  Additionally, $\bar{\I}(m)$ provides a good intuition about the qualititative nature of a given state or ensemble of states; is information generally encoded in entangled states, stored redundantly, or independently distributed among environments?

\subsection{Perfectly redundant and perfectly encoding states}
As noted previously, the primary intuition that we obtain from the $\bar{\I}(m)$ plots is that most states are ``coding'' states, but an important sub-ensemble of states are ``redundant'' states.  We are naturally led to ask whether ``perfect'' examples of each type of state exist -- that is, a state that encodes information more redundantly than any other state, or a state that hides the encoded information better than any other state.

The answer is somewhat surprising: whereas perfectly redundant states exist for any $N$, perfect coding states apparently exist only for certain $N$.  The perfectly redundant states are easy to understand; they are the generalized GHZ (and GHZ-like) states of the form:
\begin{equation}
\ket{\Psi_{\Sys\Env}} = \alpha\ket{0}_{\Sys}\Tensor_i{\ket{0}_{\Env_{i}}}
			+ \beta\ket{1}_{\Sys}\Tensor_i{\ket{1}_{\Env_{i}}}, \label{GHZLikeStates}
\end{equation}
A true GHZ state is invariant under interchange of any two subsystems; however, since mutual information is invariant under local unitaries, we only require that the states $\ket{0}_{\Env_{i}}$ and $\ket{1}_{\Env_{i}}$ be orthogonal.  Clearly, such states exist for all $N$.  Any sub-environment with $0<m<N$ has exactly $H(\Sys)$ information, but only by capturing the entire environment ($m=N$) can we obtain the full $\I = 2H(\Sys)$.  Thus, the information is stored with $N$-fold redundancy.  Note that this sort of state is slightly irregular when analyzed as in Section \ref{ProductEnvironmentSection}, since $d_i = \infty$.

A perfect coding state, on the other hand, would be one where $\bar{\I}(m)=0$ for any $m<N/2$, and $\bar{\I}(m)=\I_{\Sys\Env}$ for $m>N/2$.  An equivalent condition is the existence of two orthogonal states of $N$ qubits, each of which is maximally entangled under all possible bipartite divisions.  If such pairs of states exist, then the system states $\ket{0}$ and $\ket{1}$ can be correlated with them to produce the perfect coding state.  It is known (as detailed in \cite{ScottPRA04}) that such states only exist for $N=2,3,5,6$, and possibly for $N=7$ (for $N=6$, only a single state exists\cite{CalderbankIEEE98}).  Thus, while for large $N$ almost every state is an excellent coding state, perfect examples seem not to exist except for $N=2,3,5,(7?)$!

\subsection{Why $\bar{\I}(m)$ can be misleading}
As noted in Sections \ref{RedundancyImplicationsSection} and \ref{CaseIISection}, plots of $\bar{\I}(m)$ can sometimes be confusing or misleading.  An excellent example of this is the case of a product-environment state with a simple bimodal distribution of $d$: out of $N$ environments, $n_u$ have $d=\infty$ and $N-n_u$ have $d=0$.  This is a GHZ state with $n_u$ environments, onto which $N-n_u$ totally useless ancillas have been tacked.  As such, we note immediately that the information is stored with exactly $n_u$-fold redundancy, since any partition of the $N$ environments into $n_u$ sub-environments such that each sub-environment contains one useful environment will satisfy the condition.  Naively, we might expect that the corresponding $\bar{\I}(m)$ plot would reflect this fact, and appear like a scaled version of a GHZ-state plot.

\begin{figure}
\includegraphics[width=0.9\columnwidth]{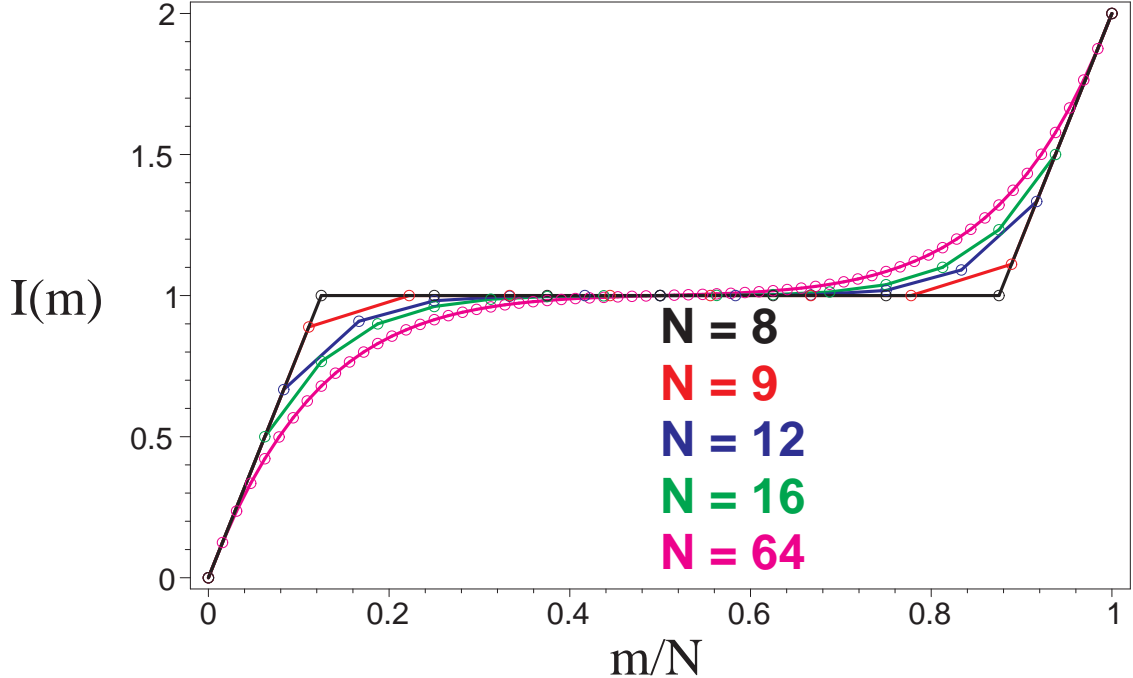}
\caption{The effect of diluting the environment.  $\bar{\I}(m)$ is plotted for environments containing $N=8,9,12,16,64$ qubits; in each case, $n_u=8$ of the qubit environments are ``useful'' ($d = \infty$), while the other $N-8$ are useless ($d=0$).  All five environments are the same from a redundancy perspective, since the extra $d=0$ qubits are irrelevant.  The $\bar{\I}(m)$ curves, however, are disproportionately distorted by the sub-environments which contain no useful information.  This dilution effect demonstrates the hazards of inferring redundancy properties from $\bar{\I}(m)$.}
\label{DilutedGHZPlot}
\end{figure}

However, as shown in Fig. \ref{DilutedGHZPlot}, diluting the environment in this way has a more dramatic effect.  Compared to the scaled GHZ-state plot, the $\bar{\I}(m)$ curves (shown for varying $N$, with $n_u$ held fixed) have been smoothed.  For sufficiently large $N$, the plots appear to indicate very little redundancy -- yet by construction, each environment has exactly the same redundancy.

The explanation is inherent in the well-known fact that $\overline{f(x)} \neq f(\bar{x})$.  If $N=8$ and $n_u=4$, then on average only $m=2$ environments
must be captured in order to obtain one useful environment (so that $\I = H(\Sys)$).  However, the probabilities of capturing (respectively) 0 and 2 useful environments do not cancel each other, since an additional useful environment adds nothing to $\I$, but failure to capture any useful environments reduces $\I$ to 0.  Thus, the unlikely event of getting no information at all is disproportionately reflected in the average over possible sub-environments.  The diluted GHZ state is the most dramatic example of this phenomenon, but it occurs in every $\bar{\I}(m)$ plot except the unimodal distribution discussed before.  Thus, while $\bar{\I}(m)$ is very useful for qualitative analysis of redundancy, it's not quantitatively reliable.

\section{Conclusions}
This paper introduces a program of analyzing \emph{how} information about a system is stored in its environment, along with some of the first insights into typical results.  While we motivate our discussion by discussing decoherence, our results are not specific to any model of decoherence, or even to decoherence processes in general.  However, the most important conclusion that we've presented is that \emph{most} states store information nonredundantly, while a certain subclass of states store information redundantly.  The evidence of the world around us indicates unambiguously that information about macroscopic objects is highly redundant in their environments, so we conclude that (a) natural processes do \emph{not} produce randomly chosen states, and (b) natural processes \emph{may} favor states of the product-environment type.

However, this is still only a conjecture \cite{ZurekADP00,ZurekRMP03}.  Such states are favored by the ultra-simplified models of quantum measurement used for toy models, but in reality both measurement and decoherence are far more complex.  We do not believe that all physical processes lead exclusively to states of the form given in Equation \ref{StateEnsemble}, but perhaps when decoherence dominates, one obtains some similar ensemble with the same properties.  The obvious next step is to analyze actual models of decoherence and quantum measurement, identify the ensembles of states produced by them, and evaluate the redundancy of information storage in those states.  We expect that this program of research will sharpen our understanding both of decoherence processes and of the classicality which they induce in the universe.

\begin{acknowledgements}
We thank ARDA for funding for this project.
\end{acknowledgements}

\bibliographystyle{apsrev}
\bibliography{/home/rbk/bib/decoherence,/home/rbk/bib/Zurek,/home/rbk/bib/quantum}

\end{document}